# Technical Report: Representing SES Cases Using Ontology

Miao Chen
1/31/2016

This is a technical report summarizing the ontology work from the FRSP project of the social-ecological informatics group at Indiana University.

**Introduction**

Socio-ecological System (SES) research studies the interaction between environment, users, and governance of environment resources. Data produced during the research cycle can be both long-tail (e.g. heterogeneous) and longitudinal data. For example, the IFRI (International Forestry Resources and Institutions) data set contains studies carried out over a period of 20 years. Given the complexity of a SES system, case studies that are accumulated over time and from different sites (e.g. site visit cases) are highly valuable in the understanding of new SES system behavior for instance. We, as a group of informatics researchers collaborating with personnel from the Workshop in Political Theory and Policy Analysis at Indiana University, are developing informatics approaches to facilitating SES scholars' research.

Here we focus on presenting our work on representing SES cases using ontology. An ontology for the SES field can help organize concepts in the field, describe resources such as data and publications using a shared vocabulary, and also facilitate data use and query for researchers. We develop a core SES ontology, which contains core concepts and resources in the field and can be used to describe actual concept and resource instances, and also a tool for contributing instances by drawing graphs, called Cmap2SES.

**SES Data**

Due to the nature of the SES studies, SES researchers collect a great amount of data in their field study. The researchers typically have a survey or questionnaire to guide the data collection and fill in the values at a given field. They can be stored in relational databases, XML files, or any customized format by the researchers. These data themselves, or extracted data, or aggregation of several data sources, are used for SES analysis. Usually researchers decide which data to use depending on their particular research questions.

Three examples of SES data sets are: Nepal Irrigation Institutions and Systems (NIIS), Common Pool Resources (CPR), and International Forestry Resources and Institutions (IFRI) databases. These data sets were originally captured in proprietary relational databases, which presented a series of risks. First, the data need long-term preservation and archiving but the proprietary relational DB is not a natural choice for data archiving because of the dependency of the data on the version of the database. Our earlier work addressed this issue by converting the IFRI relational DB, originally in Microsoft Access, to XML format by a logical mapping objects and elements from a normalized relational form into logical application-specific objects in XML (Jensen et al., 2012).

Second, the SES databases are usually isolated because researchers develop their research protocols and data storage locally. The heterogeneity among various data sets hinders larger scale SES analysis due to the difficulty of aggregating them. An overarching concept framework of the

SES domain can help with aggregating the data sets, and additionally, studies, publications, authors, and other resources in the domain. With such a concept framework positioned in the central of the network of resources of multiple types, scholars can easily navigate between resources and query relevant ones. Our early work (Chen et al. 2013) presents a data-centric approach to modeling the different types of SES resources along with a visualization interface.

We chose the IFRI data set as the sample data for applying the tools here because we have access to the data. The IFRI data results from long-term data collection at various forests around the world over the last 20 more years, first starting in 1993 (IFRI, 2015). Field researchers collect local forest data according to the IFRI survey questionnaire and then upload the data to a central database in Microsoft Access. This data has the typical characteristics of SES data: ranging a long-term life span and collected by researchers at different places. We obtained a snapshot of the IFRI relational database and use it as example throughout the study to showcase how we address the ontology representation issues of SES cases.

The IFRI database design roughly follows the concept model of IFRI data collection shown below. Typically a form is mapped to a table in the database and the relations were translated to foreign keys. For example, form F (Forest form) is a table, form P (Forest Plot form) is another table, and a forest can have multiple plots through using foreign key.

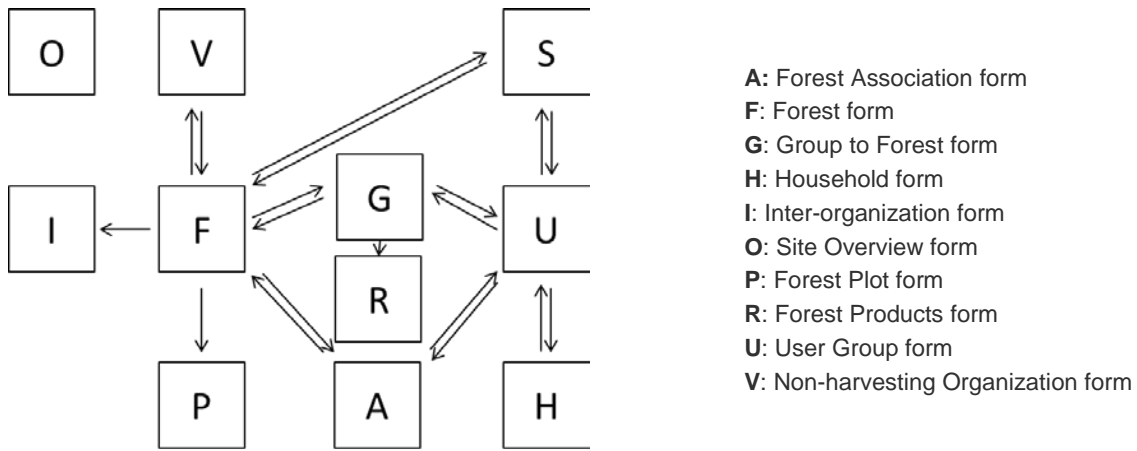

**A:** Forest Association form
**F**: Forest form
**G**: Group to Forest form
**H**: Household form
**I**: Inter-organization form
**O**: Site Overview form
**P**: Forest Plot form
**R**: Forest Products form
**U**: User Group form
**V**: Non-harvesting Organization form

Figure 1. Conceptual model of IFRI data collection (available from http://www.ifriresearch.net/wp-content/uploads/2012/09/MethodsImage.png)

**Representation Challenges**

Researchers often develop their research protocol and data representation style locally, and thus SES cases currently do not share the same representation style. They may be stored in plain text files, XML files based on some metadata standard, or relational database. The representation style of one researcher may vary greatly from another. Though researchers who collect the data can conveniently use them, heterogeneity across SES data sets can bring difficulties when aggregating data sets of different representations to analyze SES systems at a larger scale.

More specifically, representation heterogeneity of SES data occurs in two aspects: heterogeneity in description format and heterogeneity in vocabulary. The former is about representation style such as XML, text files, metadata, ontology, and the latter is about a shared set of concepts for describing the SES system. Both these two heterogeneity issues can be resolved by using ontology representation, which can offer a semantic based representation style and a shared vocabulary of rich semantics. For building the ontology, we use semantic web languages such as OWL and RDF, which are standard languages to describe concepts and resources on the semantic web.

The SES data aggregation suffers from heterogeneity. When scholars use the data they collect locally, data mapping does not appear as a prominent problem; however in the case of aggregating use cases from different studies, an activity not uncommon in SES research, mapping data fields of different data sets present a huge challenge to researchers. Without a shared vocabulary of concepts of the field, it is very difficult to link between different data sets; also without a shared description format of the data, it is challenging to understand the semantics of the description and further establish connections between data fields. Therefore ontology is a feasible solution to the heterogeneity issue because it can contain a shared vocabulary of domain concepts as well as provide a semantic description of the resources.

**The SES Framework as a Shared Vocabulary for the Ontology**

The SES field, as any field, needs a shared vocabulary to describe resources and establish connections between them. A conceptual framework of the field can serve as this vocabulary. Ostrom (2009) proposed a seminal framework for conducting analysis on SES systems, followed by several revisions (e.g. McGinnis & Ostrom, 2014). An SES scholar can adopt the SES framework to build an analysis framework for their particular SES case, e.g. a particular forest or watershed, because it provides both vocabulary and structure for the analysis. The application of an SES framework to a particular case is discussed in more details in the following section. An SES case can then be added to the ontology after it is analyzed in the SES framework.

In addition to the analysis framework, it also provides a hierarchy of concepts, which is an ideal candidate for serving as a shared vocabulary given the fact that this framework is widely cited. We make use of these hierarchical concepts as the shared vocabulary and include them in the ontology. With concepts in Ostrom's (2009) and McGinnis & Ostrom's (2014) frameworks serving as the concept hub of the ontology, the various SES data and resources can be described according to the shared vocabulary. That is to say, the SES case data are first described using the concepts, and are then linked together via the SES concepts.

Below are two SES framework examples, presented in Ostrom (2009) and McGinnis and Ostrom (2014) respectively. In each figure, the left diagram shows an SES framework graphically including first-tier concepts and interactions between concepts, and the right diagram shows more details with second-tier concepts in a hierarchical tabular format.

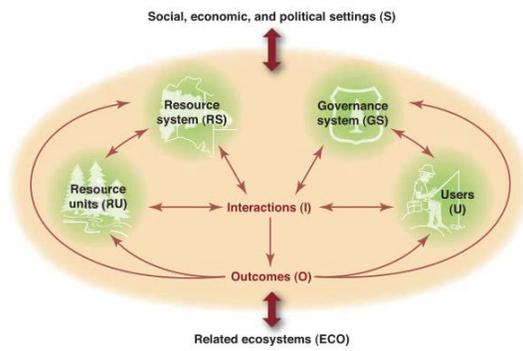

Figure 2. The SES framework in Ostrom (2009).

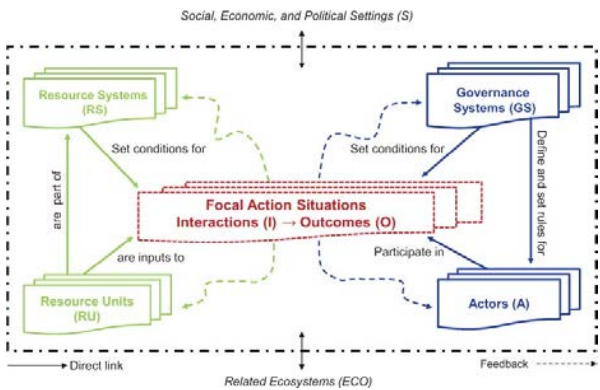

Figure 3. The SES framework in McGinnis & Ostrom (2014).

The above two graphs also reflect the evolvement of the SES framework over time: for example, the concept *Users* in Ostrom (2009) is changed to *Actors* in McGinnis & Ostrom (2014); and explicit relations between concepts are added to the McGinnis & Ostrom (2014) SES graph, e.g. Resource Systems *set conditions for* Focal Action Situations, etc.

Scholars may use different versions of SES framework for the analysis of the SES systems of their interest, especially before the advent of the updated models. That means to describe a data set accurately we need to distinguish between a same named concept in frameworks of different versions. For example, the concept "Resource Systems" in Ostrom (2009) is a different concept entity from the "Resource Systems" in McGinnis & Ostrom (2014), because although they share the same name they are two different concept entities contained in different frameworks. Moreover, the meaning of a concept changes over time so it is necessary to distinguish concepts of a same name but from different times and frameworks.

The concepts and relations in the SES framework provide a good vocabulary source to construct the SES ontology. In this study, we mainly use the actual concepts in SES as instances of the

Concept class, as well as the hierarchical relations between concepts to further help construct SES instances in the ontology.

**SES Cases** (a.k.a. SES Instances)

An SES case can be defined as a case about a particular SES system or a suite of SES systems. It can be a site visit to a forest, or a study of a water resource, or a study of the forest system in Indonesian. For example, In the IFRI database, each site visit can be considered as an SES case. In the ontology language, an SES case if frequently referred to as an SES instance. Therefore we use *SES case* and *SES instance* interchangeably here.

Description of an SES instance can be distilled from published papers using SES frameworks. Such papers are usually resulted from analysis of one or multiple site visits under an SES framework. For instance, Cox's (2014) study applies the SES framework on Tao's village irrigation system. It contextualizes the SES concepts in the Tao's village, for example, by extending the SES framework by additional tiers of concepts and specifying meaning of original SES concepts in this particular context. The figure below presents the contextualized SES framework as used in Cox's (2014) study:

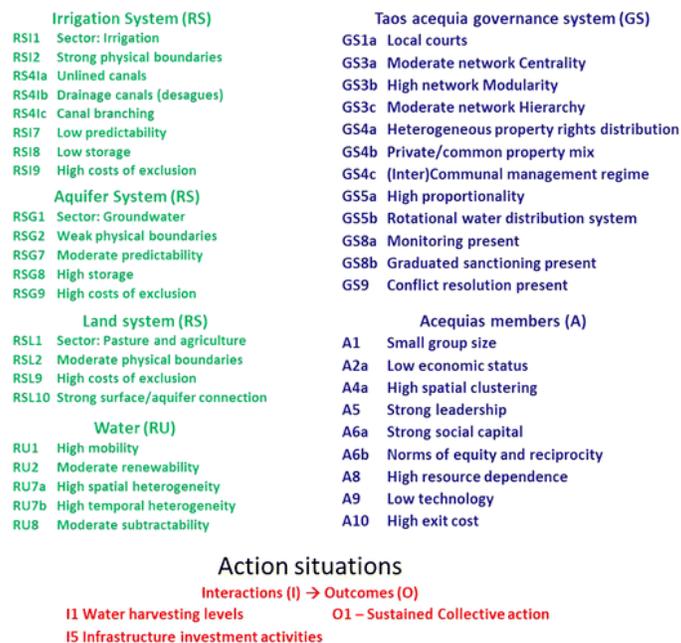

Figure 4. The SES framework contextualized in Cox's (2014) study.

The above figure shows how an actual SES case can be analyzed within an SES framework. That also means the original framework needs to be contextualized based on the conditions of the actual case. For example, Cox's (2014) study has Irrigation System, Aquifer System, Land System, which all map to the Resource System in the original framework, meaning the Resource System are contextualized to the three systems in this context.

In some publications, the researcher may not explicitly state the SES description of their cases. Therefore some more manual work is needed to extract the description using the SES framework. This kind of work needs intensive human understanding of the study and the SES field in general to extract the SES description, but once the needed information for the ontology is extracted it can bring great benefits to researchers.

The SES cases, once described by one of the SES framework, are ready to be added to the SES ontology. The adding of the cases can be done by writing the case in an RDF/XML file with referring to the SES core ontology, and alternatively can also be achieved by drawing a graph using the Cmap2SES tool we developed. For researchers who are unfamiliar with the RDF semantic web language, using the Cmap2SES tool to describe a case is a nice option. These added cases enrich the ontology with actual cases such that researchers can query SES cases using free-text keywords or structured query.

**SES Core Ontology**

We have developed an abstract ontology to describe main resources and properties in the SES field, called "SES-core". "SES-core" can be used as a template for researchers to contribute their SES cases as instances to the SES ontology. We use the semantic web languages OWL and RDF for describing classes and instances in the ontology. The full SES-core ontology is accessible at https://github.com/Data-to-Insight-Center/socioeco-Cmap2SES/blob/master/onto/sescore

Main classes. The SES-core abstract ontology contains the following primary classes, as illustrated in Figure 5: Concept, GlobalConcept, Local Concept, Framework, ConceptGraph, Study, Publication, Author. They can be used to describe SES framework and individual SES cases. For expressing concept hierarchy in the SES framework and SES cases, we make use of the readily available SKOS vocabulary, which provides skos:broader and skos:narrower predicates, as well as some predicates created specifically for this ontology, such as refers_to.

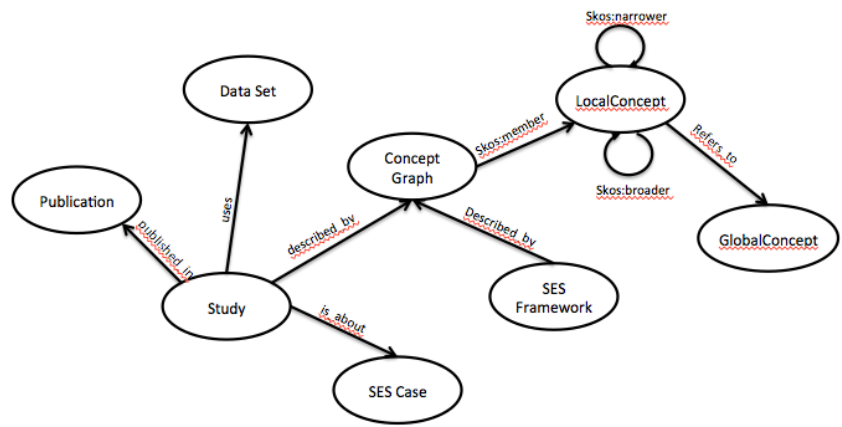

Figure 5. Main classes and properties of the SES-core ontology.

***Global Concept* vs. *Local Concept***. Global Concept and Local Concept are subclasses of the Concept class. Concepts contained in an actual SES instance or Framework, such as the ones in Ostrom(2009) and Cox(2011) are considered as LocalConcept instances referring to a GlobalConcept. For example, the concept "Resource System" is a top-tier category in SES framework in Ostrom(2009). To represent a resource system, we create a local concept for the "Resource System" as described by Ostrom(2009) and then refer it to its global version, i.e. the

ResourceSystem concept. The triple statement can be expressed as: *ResourceSystem_Ostrom2009 refers_to ResourceSystem*.

**Adding a Case to the SES Ontology: An Example using Cox's (2014) Study**

We show an example of adding a case to the SES ontology using Cox's (2014) study on Acequia irrigation system as an example. Comparing the SES framework in McGinnis and Ostrom (2014) and the one in Cox (2014), we can find in Cox (2014), the Resource System (RS) in McGinnis and Ostrom (2014) is contextualized to Irrigation System, Aquifer System, and Land System, Resource Unit (RU) is contextualized to Water.

**Cmap2SES**

It is clear that in principle, adopting an ontology-based representation scheme brings the benefits of a semantic representation. However, the ability to realize these benefits in practice may be limited because such representations may not easily be generated by domain specialists who are novices at ontology generation and languages such as OWL and RDF. It is a non-trivial task for scholars, especially those new to ontology languages, to populate a SES instance compatible with the SES ontology. The prerequisites include both understanding of ontology languages and the specific structure of the SES ontology.

The graph drawn in a concept map, however, is different from the final statements to be added to the ontology. Users do not need to explicitly specify the category of a node, e.g. a node labeled as "Resource System", is automatically assigned to the Concept category. However, this information is critical for an ontology as we need to claim the class information of a node in the SES case for completeness and further inference.

In order to reduce the representational barrier to creating SES cases, we have developed a concept map template to help users easily create an SES case by drawing a graph. The graph is later converted to an RDF snippet for the SES ontology. We use the Cmap Ontology Edition (COE) tool ("Cmap COE", 2015), a graphical interface that enables users to draw ontology elements and convert the resulting concept map to an OWL/RDF file. Cmap ("Cmap", 2015) is a graphical tool for constructing concept maps for knowledge representation and organization, and Cmap COE a tool for constructing ontologies graphically based on Cmap. In Cmap COE, users draw ontology entities as nodes, and draw entity properties, which connect between entities, as edges.

Cmap COE allows users to import pre-defined templates in a graph such that they can develop their own graph based on the templates. Our template for SES cases, called the SES Cmap Template, delineates important classes for an SES instance as nodes, and important properties as edges in a graph. Figure X shows the template, called "SES template". It can be used as basis for drawing concrete SES cases. The template contains these abstract statements:
- sescore:SES_Study sescore:described_by sescore:ConceptGraph (an SES study is described by a concept graph)
- sescore:ConceptGraph skos:member sescore:Concept (a concept graph has member of concept/s)
- sescore:Concept skos:member SubConcept (a concept has subconcept) (note: the SubConcept is actually a Concept too but we label it as SubConcept too because that can help users know they should fill in a child concept of a particular concept)

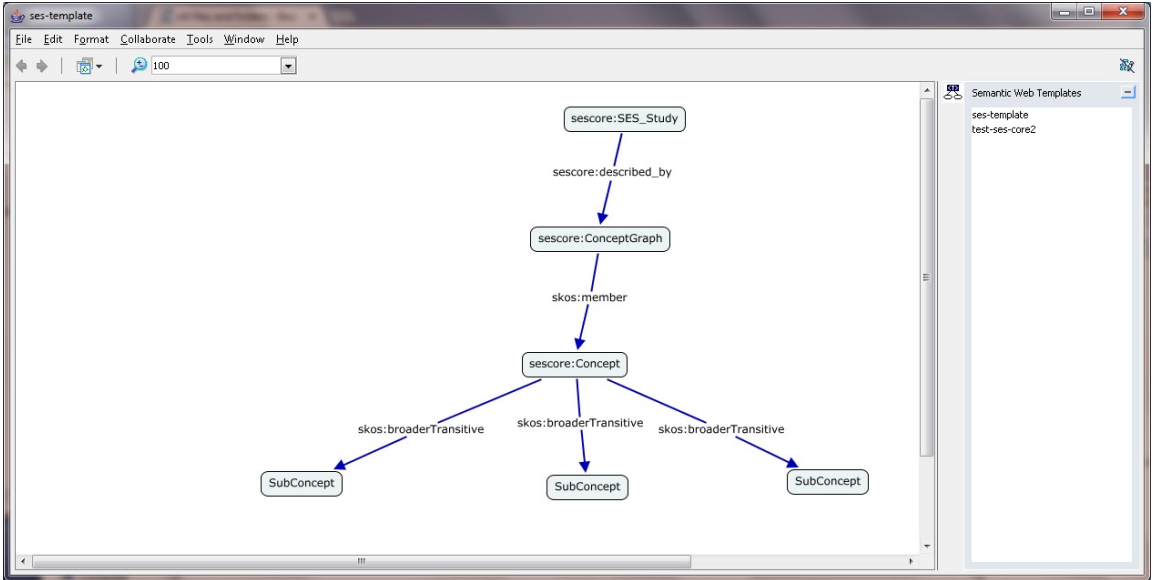
Figure 6. SES template in Cmap COE.

Users can import the above template when creating their own Cmap about the SES case, and fill in corresponding values based on the template. For example, fill in a real concept, "ResourceSystem" at the place of the "sescore:Concept". To create an SES case in Cmap, first, users need to create a new Cmap in the Cmap COE interface. Then on the right pane click on the graphic button for templates and all the available templates will show up.

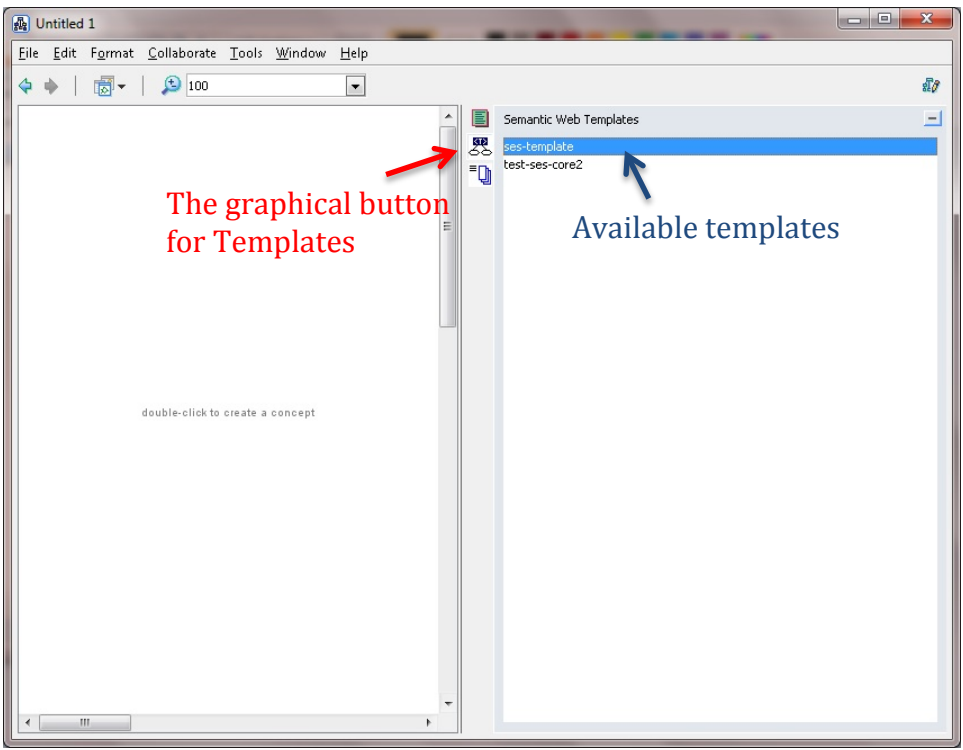

Figure 7. "ses-template" is shown on the right pane after clicking on the graphic button for templates.

Users can create their SES case by dragging the "ses-template" template on the right pane to the Cmap editing interface on the left and then filling in more elements on their own, as shown below. The template is graphically an abridged version of the SES-core ontology shown in Figure 8.

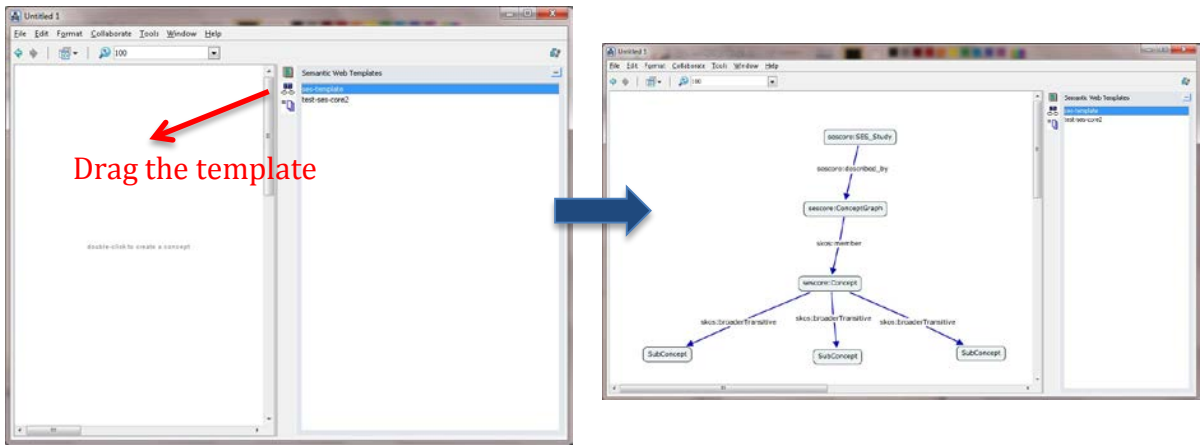

Figure 8. Dragging template on the Cmap editing interface.

By importing the "SES-template" in their Cmap, the users obtain a new Cmap and they need to further edit it to reflect the situations in their SES case. Uses can also edit it to an SES framework, such as the frameworks in Ostrom (2007), Ostrom (2009), and McGinnis and Ostrom (2014). Regardless their goal being an SES case or SES framework, the approach in Cmap COE is the same, by using the SES-template and adding more details to it.

We present an example of drawing the SES framework mentioned in Ostrom (2007) in Cmap COE. After important the template, we perform these actions in the Cmap COE interface:
- rename "sescore:SES_Study" to "Ostrom2007_SESFramework" to reflect this is an actual framework mentioned in a publication;
- rename "sescore:ConceptGraph" to "ConceptGraph_Ostrom" to reflect it is an instance of the "sescore:ConceptGraph" class;
- Add more Concept nodes whose upper arrow starts from "ConceptGraph_Ostrom2007"; these nodes are the first-tier categories in Ostrom's (2007) framework and are named "ResourceSystem_Ostrom2007", "GovernanceSystem_Ostrom2007", "User_Ostrom2007" respectively.
- Since each first-tier categories have child categories, we reflect it by adding more child nodes to them. For example, we add two child concepts, "RS9_Location_Ostrom2007" and "RS1_Sector_Ostrom2007", as subconcepts of "ResourceSystem_Ostrom2007" by using the *skos:narrower* predicate as the edge between the parent and the child.

Below shows the resulted Cmap, reflecting part of the Ostrom (2007) framework:

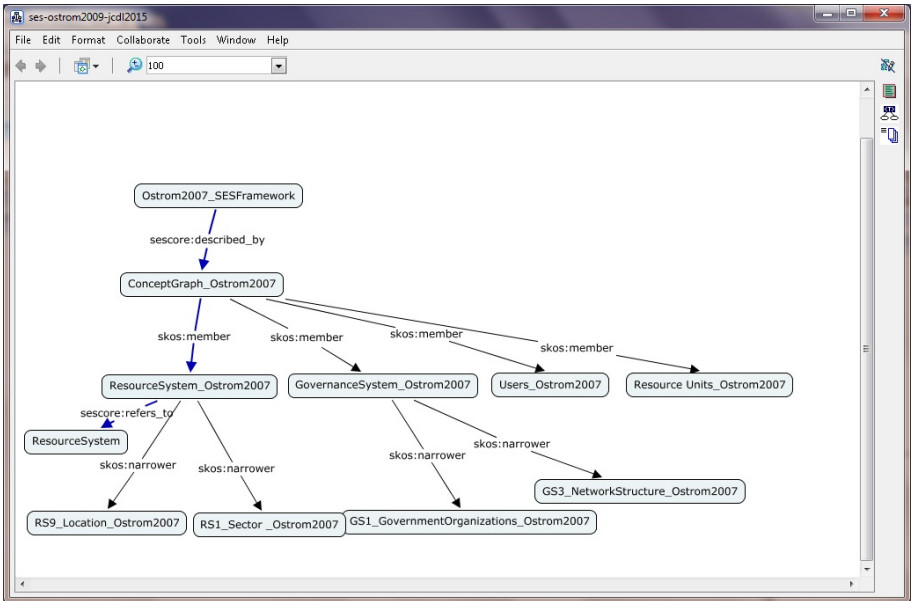

Figure 9. The Cmap drawing for the Ostrom's (2007) framework (part of it).

After finishing the concept graph, users can export it to an OWL file in the Cmap COE interface and submit it to our web server using the Cmap2SES service. The Cmap COE tool provides a function to export a Cmap graph to an OWL file, as shown below.

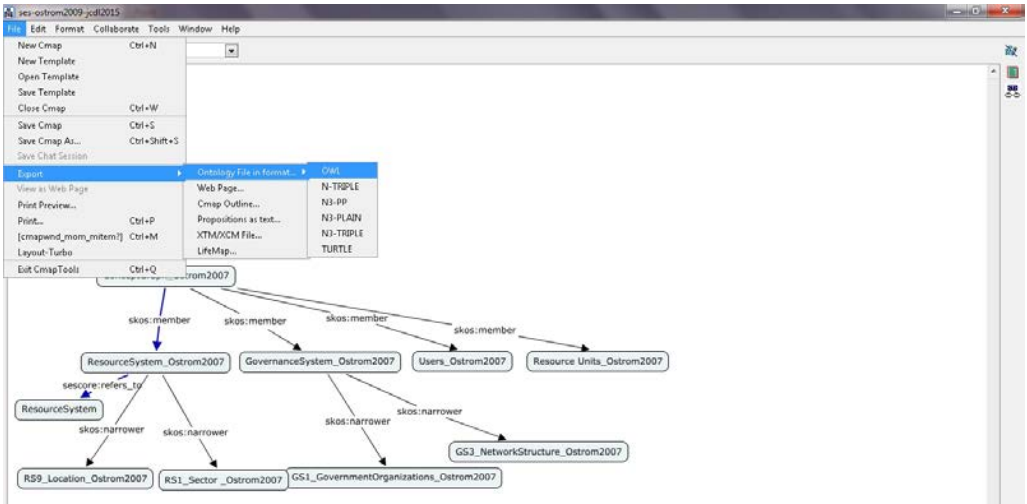

Figure 10. Export a Cmap drawing to an OWL file, in Cmap COE.

This automatically generated OWL is a valid OWL file in terms of its syntax however it is incompatible with the SES-core ontology due to: 1) Cmap COE considers any node in the graph as an ontology class, which is not always true in our case.  For example, "Resource System", "Governance System" are actually instances of the Concept class however they are made as Resource System class and Governance System class if without proper intervention; 2) more statements need to be added to the OWL because users typically do not express obvious knowledge in the graph. For example, "Resource System" itself is an instance of the Concept class but users are not expected to add this obvious knowledge, so we choose to add this statement at our backend service once users submit their OWL files generated from Cmap COE.

Therefore to ensure compatibility between the Cmap COE generated OWL files and the SES-core ontology, we have developed a web service, called "Cmap2SES", in Apache Tomcat 7 to transform the uploaded OWL files into a SES-core compatible form. The web service addresses the above two compatibility issues by explicitly stating the nodes in the concept graph as instances of the LocalConcept class and adding assertions for some knowledge obvious to humans and likely to be implicit in the human-defined representation: e.g. declaring concepts as instances of LocalConcept, referring the local concepts to their global concept versions, and creating their global concept instances if they are not in the GlobalConcept yet.

**Related Work**

Research data sets can be described using a variety of vocabularies and formats, ranging from plain text to formats with metadata characterizations, and ontology-based representations. Standard or shared vocabularies facilitate understanding by community users such that scholars can understand the data set description by referring to the vocabulary definition. These vocabularies often contain classes for important resource types in the domain as well as predicates and properties to describe relations between resources. Examples include Dublin Core (DC), for basic resource description, and the Simple Knowledge Organization System (SKOS), for knowledge organization system description. For the ecology domain, the Ecological Metadata Language (EML) vocabulary is available for describing data sets.

However, the descriptions of many legacy data sets do not follow a standard or shared vocabulary. These data sets were collected relatively early (e.g. the IFRI data used in our study was collected in the 1990's) and data protocols were usually locally developed by researchers. Such data descriptions can serve scholars' data needs well if they only need to use a local subset, for example, as in studies where subsets of the IFRI data are used (Persha, Agrawal, & Chhatre, 2011; Poteete & Ostrom, 2004). However, for larger scale analysis aggregating data from different sources, the usefulness of a locally developed data description without shared semantics is limited because it is difficult to align various data sets without intensive manual work. If data fields of different data sets can be aligned by using shared semantics, then data aggregation and large-scale SES analysis becomes more possible.

Information science and SES researchers alike have explored methods for assigning shared semantics to the SES data, both for ongoing data collection and for legacy data. One approach is to develop a central database to describe SES cases in entity-relation fashion, as demonstrated in the SESMAD and SES Library projects (Cox, 2014b; "SES Library"). Another is to make use of the SES frameworks proposed by domain scholars. Such frameworks identify social and ecological variables, as in the seminal SES framework proposed in (Ostrom 2007; 2009; McGinnis & Ostrom, 2014). These variables can be used to label data sets and thus heterogeneous social and ecological data sets can be linked for SES analysis (Kittinger, Finkbeiner, Glazier, Crowder, 2012). An overarching framework across social and ecological domains can greatly facilitate the integration of various data sets (Altaweel, Alessa, Kliskey, & Bone, 2010).

We try to complement the existing work in the field by building a workflow with relevant components integrated. Our current work is also built upon our earlier work in archiving SES legacy data via constructing logical objects (Jensen et al., 2012). Our current effort supports representation of SES knowledge using an ontology and capture of SES cases via domain experts drawing concept models, as well as graph modeling and mining of the fine-grained cases within a data set. This work also relates to artificial intelligence research on knowledge capture, and in particular, to the use of concept mapping as a precursor to ontology generation.

# References

Altaweel, M. R., Alessa, L. N., Kliskey, A., & Bone, C. (2010). A framework to structure agent-based modeling data for social-ecological systems. *Structure and Dynamics*, *4*(1).

Cmap. http://cmap.ihmc.us/

Cmap COE. http://www.ihmc.us/groups/coe/

Cox, M. (2014a). Applying a social-ecological system framework to the study of the Taos Valley irrigation system. *Human Ecology*, *42*(2), 311-324.

Cox, M. (2014b). Understanding large social-ecological systems: introducing the SESMAD project. *International Journal of the Commons*, *8*(2), 265-276.

Dublin Core Metadata Initiative. http://dublincore.org/

EML (Ecological Metadata Language). https://knb.ecoinformatics.org/#tools/eml

International Forestry Resource and Institutions (IFRI). http://www.ifriresearch.net/

Jensen, S., Plale, B., Liu, X., Chen, M., Leake, D., & England, J. (2012, October). Generalized representation and mapping for social-ecological data: Freeing data from the database. In *IEEE 8th Int'l Conf on E-Science (e-Science),* (pp. 1-8). IEEE.

Kittinger, J. N., Finkbeiner, E. M., Glazier, E. W., & Crowder, L. B. (2012). Human dimensions of coral reef social-ecological systems. *Ecology and Society*, *17*(4), 17.

McGinnis, M. D., and E. Ostrom. 2014. Social-ecological system framework: initial changes and continuing challenges. *Ecology and Society* **19**(2): 30. http://dx.doi.org/10.5751/ES-06387-190230

Ostrom, E. 2007. A diagnostic approach for going beyond panaceas. *Proceedings of the National Academy of Sciences* 104(39):15181-15187. http://dx.doi.org/10.1073/pnas.0702288104

Ostrom, E. 2009. A general framework for analyzing sustainability of social-ecological systems. *Science*325:419-422. http://dx.doi.org/10.1126/science.1172133

Persha, L., Agrawal, A., & Chhatre, A. (2011). Social and ecological synergy: local rulemaking, forest livelihoods, and biodiversity conservation. *Science*, *331*(6024), 1606-1608.

Poteete, A. R., & Ostrom, E. (2004). Heterogeneity, group size and collective action: The role of institutions in forest management. *Development and change*, *35*(3), 435-461.

SES Library. https://seslibrary.asu.edu/seslibrary/

SKOS (Simple Knowledge Organization System). http://www.w3.org/2004/02/skos/